# Non-Local Macroeconomic Transactions and Credits-Loans Surface-Like Waves

Victor Olkhov

TVEL, Kashirskoe sh. 49, Moscow, 115409, Russia,

victor.olkhov@gmail.com

Abstract

This paper describes surface-like waves of macroeconomic Credits-Loans transactions on economic space. We use agent's risk ratings as their coordinates and describe evolution of macro variables by transactions between agents. Aggregations of agent's variables with risk coordinates *x* on economic space define macro variables as function of *x*. Aggregations of transactions between agents at point *x* and *y* determine functions of two variables (*x,y*) on economic space. As example we study Credits transactions provided from agents at point *x* to agents at point *y* and thus amount of Loans received by agents at point *y* from agents at point *x* at moment *t* during time term *dt*. We model evolution of macro transactions by hydrodynamic-like equations. Agents fill macro domain on economic space that is bounded by minimum risk ratings of most secure and maximum risk ratings of most risky agents. Economic and financial shocks can disturb steady borders of macro domain and cause perturbations of transactions. Such disturbances can generate waves that can propagate along risk borders *alike to* surface waves in fluids. As example, we describe simple model interactions between two transactions by hydrodynamic like equations in a closed form. We introduce notions of "*macro accelerations*" and their potentials that establish steady state distributions of transactions on economic space. For this model in linear approximation we describe surface-like waves and show that perturbations induced by surface-like waves can exponentially grow up inside macro domain and induce macro instabilities in a low risk area. Description of possible steady state distributions of transactions and surface-like waves on economic space might be important for macro modeling and policy-making.

JEL: C00, E00, F00, G00[1]

---

[1] This research did not receive any specific grant, financial support or assistance from TVEL or funding agencies in the public, commercial, or not-for-profit sectors and was performed on my own account only.

# 1. Introduction

Description of time fluctuations of economic and financial variables as Demand and Supply, Investment and Profits and modeling volatility of Asset prices and Returns defines core problems of macroeconomics. Diversity of economic environment that may cause fluctuations and variety of macro variables that follows such fluctuations make this problem very difficult. Description of macro fluctuations and their consequences are presented in numerous recent papers [1-4]. In our work we regard small slice of these tough problems and present framework that may describe origin of fluctuations of macro variables. As usual macro variables are treated as functions of time and their fluctuations as oscillations in time. However time fluctuations may reflect action of various hidden wave processes. Description of any waves requires certain space where such waves can propagate. In [5-9] we introduce economic space as ground for economic and financial modeling and describe different macro waves. In present work we study one else type of macro waves and explain how they may cause time fluctuations of economic and financial variables.

All macro variables are defined by corresponding variables of separate economic agents those compose macro system. For example, macro Capital or Assets are defined by aggregative values of Capital or Assets of all economic agents. Let assume that macroeconomics is composed by set of agents each described by numerous financial and economic variables. Let propose that it is possible estimate risk ratings $x$ for each agent and let use agent's risk ratings as their coordinates $x$ on economic space. That allows attribute variables of each agent by same coordinates $x$ on economic space [5-9]. Hence variables of agents as Assets or Capital can be defined as functions of time $t$ and coordinates $x$ on economic space. Variables of agents with same coordinates $x$ define macro variables as functions of time and coordinates. For example sum (without doubling) of Assets or Capital of all agents with coordinates $x$ define macro Assets or Capital as functions of time $t$ and coordinates $x$ on economic space. Sum or integral of macro Assets by coordinates $x$ over economic space defines Assets of entire macroeconomics as function of time only. We propose that waves of macro variables on economic space may cause time fluctuations of macro variables. For example waves of macro Assets or Capital on economic space may cause time fluctuations of Assets or Capital of entire economics. Thus description of time fluctuations of variables requires development of wave models of macro variables on economic space. We study this problem here.

Macro variables at point $x$ on economic space are determined as aggregates of corresponding financial variables of agents in the same domain. Evolution of agent's variables defines evolution of macro variables. For example evolution of macro Assets is determined by change of agent's Assets. Agent's variables are changed via transactions between agents. For example transactions between agent $A$ and agent $B$ can describe change of Credits from $A$ to $B$, growth or decline of Profits of $A$ received from $B$, increase or fall of Consumption of $A$ that Supplied by *B* and etc. Changes of agent's variables define change of macro variables. Thus transactions between agents define dynamics of macro variables. Fluctuations of transactions between agents *A* and *B* induce fluctuations of corresponding macro variables. Description of macro variables is extremely complex and requires different approximations. In [5-7, 9] we described macro model with transactions in a *local* approximation that proposes that transactions occur between agents with nearly same coordinates *x*. In [8] and in this paper we study macro model that describes transactions between agents at any points *x* and *y* on economic space. For example economic agent *A* with particular risk rating *x* can Buy and Sell Assets and Goods or provide Credits to agent *B* with any risk rating *y*. Such non-local transactions define "action-at-a-distance" between agents *A* and *B* on economic space. Modeling transactions between agents at different points on economic space describe complex interrelations between macro variables. Aggregation of all transactions from agent *A* at *x* to all agents at point *y* defines transaction from agent *A* at *x* to point *y*. Aggregation of all transactions from all agents at point *x* to all agents at point *y* define macro transactions between points *x* and *y* as function of time *t* and two variables (*x,y*).

S*imilar* models of transactions between agents are well known in economics as Leontief's framework [10-13]. The substantial difference between our approach and Leontief's input-output model is follows. Leontief allocates economic agents by industries. He describes transactions by resources between industries in such a way that output from one industry becomes input of another one. We replace Leontief's allocation of agents by industries with allocation of agents by their risk ratings as coordinates on economic space. That allows replace transactions between industries with transactions between different points of economic space. Importance of such "small" modifications is this: Leontief's partition of entire economics by industries does not define any linear space. Our allocations of agents by their risk ratings as coordinates on economic space describe macro transactions on a linear economic space. Such "simple" replacements of allocations by industries with distributions of agents by their risk ratings on economic space permits describe macroeconomics and finance in a completely different manner.

Risk ratings of agents are reduced by minimum and maximum grades that reflect most secure and most risky agents. Thus all agents of macro system fill certain domain on economic space. Borders of such domain describe most secure and most risky macro areas on economic space. Macro transactions between points *x* and *y* as function of two variables (*x,y*) where *x* or *y* are near or belong to macro borders depend on variations of agent's variables. Perturbations of macro domain borders by economic and financial shocks cause disturbances of macro transactions. Such perturbations of macro transactions with one coordinate near macro border can generate waves that are *alike to* surface waves in fluids [14,15]. Nature of macro waves and nature of surface waves in fluids are absolutely, completely different and we use only *wording* parallels between them. Surface-like waves of macro transactions disturbances may propagate along macro border or inside macro domain. Surface-like waves of transactions may induce waves of macro variables and cause time fluctuations of variables of entire macroeconomics.

This paper describes macroeconomics with *non-local* transactions between agents at points *x* and *y* on economic space and derive surface-like wave equations. As example we study Credits-Loans transactions and describe their surface-like waves on risk borders of macro domain. We present simple surface-like waves of Credits-Loans transactions and study their influence on evolution of macro variables as functions of coordinates on economic space and on time fluctuations of macro variables. Even simple Credits-Loans surface-like waves admit exponential amplification inside macro domain and that may cause macro instabilities at secure macro area with low risk coordinates.

The rest of the paper is follows. In Section 2 we argue model setup and repeat definitions of economic space, macro variables and macro transactions between two points on economic space. In Section 3 we describe dynamics of macro transactions by hydrodynamic-like equations. In Section 4 we study simple model of interactions between two macro transactions on macro domain and derive surface-like wave equations. We study simple wave solutions and present examples that describe exponential amplification of small border perturbations inside macro domain. That may model impact of small disturbances of macro transactions on sustainability of macroeconomics in secure and low risk area. Conclusions are in Section 5.

## 2. Model Setup

Economic space notion and definition of macro variables on economic space were presented in [5-9]. We refer these papers for all further details. For convenience we briefly argue these notions below.

### *2.1. Economic space and macro variables*

Let regard macroeconomics as an ensemble of economic agents those perform various transactions with one another. Let assume that it is possible to estimate risk ratings of all agents; for huge corporations and for small firms or households. Risk ratings take values of risk grades and we propose treat grades of single risk as points of one-dimensional space and simultaneous risk assessments of *n* different risks as measurements of agent's coordinates on *n*-dimensional space. Let propose, that risk assessments methodologies can be generalized in such a way that risk grades can take continuous values and define space $R$. Thus risk grades of *n* different risks establish $R^n$. Let define economic space as space that allows map agents by their risk ratings as space coordinates. Below we study macroeconomics and finance under action on *n* risks on economic space $R^n$.

Agent-based models are usually associated with equilibrium, behavioral or decisions making [16-18]. We don't argue any general equilibrium models or behavioral issues. Let regard economic agents in a completely different manner and replace question: "Why economic agents take certain decisions?" with a different one: "How evolution of agent's variables describe macroeconomics and finance?"

Usage or risk ratings as agent's coordinates on economic space allows establish *wording* parallels between macroeconomics and multi-particle systems in physics keeping in mind that their nature is absolutely different. Nevertheless *language* similarities help develop macro model on economic space. For brevity let call economic agents of macro system as economic particles or e-particles and economic space as e-space. Each e-particle has many financial and economic variables like Credits and Loans, Assets and Debts, and etc. Macroeconomics and finance as ensemble of e-particles on e-space can be treated *alike to* certain "*gas*". Economic and financial variables of separate e-particles define macro variables of such "*gas*" *alike to* mass of separate particles define mass density of gas or fluid in physics. For simplicity let assume that all e-particles are "independent" and sum of any extensive (additive) variable of any *k* e-particle equals same variable of entire group of these *k* e-particles. That assumption allows collect values of extensive variables of different e-particles without fear of double accounting. If all e-particles of macroeconomics are "independent" then sum of Assets $a(t,x)$ of all e-particles with coordinates $x$ defines macro Assets $A(t,x)$ as function of $x$ on e-space. Integral of macro Assets $A(t,x)$ by $dx$ over e-space defines Assets $A(t)$ of entire macroeconomics as function of time *t* only. Such simple considerations allow perform transition from description of evolution of variables of separate e-particles to problem of evolution of macro variables as functions of time and *coordinates* on e-space. This transition of macro description has parallels to transition from kinetic description of multi-particle system in physics that takes into account granularity of separate particles to continuous media or hydrodynamic approximation. To evaluate such parallels let mention that due to random nature of risks and action of economic and financial processes on e-particles variables their aggregations near point $x$ should be random. For example sum of Assets of e-particles with coordinates near $x$ should fluctuate due to fluctuations of Assets of separate e-particles. To obtain macro variables as a regular functions of time *t* and coordinates $x$ on e-space let apply averaging procedure.

In a more formal way [9] let assume that there are $N(x)$ e-particles at point $x$. Let state that velocities of e-particles at point $x$ equal $v=(v_1,… v_{N(x)})$. Each e-particle has *l* extensive variables ($u_1,...u_l$). Let assume that values of variables equal $u=(u_{1i},...u_{li})$, $i=1,..N(x)$. Each extensive variable $u_j$ at point $x$ defines macro variable $U_j$ as sum of variables $u_{ji}$ of $N(x)$ e-particles at point $x$

$$U_j = \sum_i u_{ji}; \quad j = 1,..l; \quad i = 1,...N(x)$$

To describe motion of variable $U_j$ let establish additive variable alike to impulse in physics. For e-particle $i$ let define impulses $p_{ji}$ as product of extensive variable $u_j$ that takes value $u_{ji}$ and its velocity $\boldsymbol{v_i}$:

$$p_{ji} = u_{ji}\boldsymbol{v_i} \qquad (1.1)$$

For example if Assets $a$ of e-particle $i$ take value $a_i$ and velocity of e-particle $i$ equals $\boldsymbol{v_i}$ then impulse $p_{ai}$ of Assets of e-particle $i$ equals $p_{ai} = a_i\boldsymbol{v_i}$. Thus if e-particle has $l$ extensive variables ($u_1,...u_l$) and velocity $\boldsymbol{v}$ then it has $l$ impulses $(p_1,p_2,..p_l)=(u_1\boldsymbol{v},...u_l\boldsymbol{v})$. Let define impulse $\boldsymbol{P_j}$ of variable $U_j$ as

$$\boldsymbol{P_j} = \sum_i u_{ji}\boldsymbol{v_i}; \quad j = 1,..l; \quad i = 1,...N(x) \qquad (1.2)$$

Reasons to introduce *"economic"* impulse are follows. Velocities are non-additive and sum of agent's velocities don't define velocity of any group of agents. Impulses (1.1) are additive and thus allow define impulse $\boldsymbol{P_j}$ of variable $U_j$ by Eq.(1.2). As we show below (Eq.(2.3)) that help define velocity of variable $U_j$. Let introduce distribution function $f=f(t,x;U_1,..U_l, \boldsymbol{P_1},..\boldsymbol{P_l})$ that determine probability to observe variables $U_j$ and impulses $\boldsymbol{P_j}$ at point $x$ at time $t$. $U_j$ and $\boldsymbol{P_j}$ are defined by corresponding values of e-particles that have coordinates $x$ at time $t$. They take random values at point $x$ due to random motion of e-particles on e-space. Averaging of $U_j$ and $\boldsymbol{P_j}$ via distribution function $f$ allows establish transition from approximation that takes into account variables of separate e-particles to continuous *"media"* or hydrodynamic-like approximation of macroeconomics that neglect e-particles granularity and describe macro variables as regular functions of time and coordinates on e-space. Let define density functions $U_j(t,x)$ as

$$U_j(t,\boldsymbol{x}) = \int U_j\, f(t,\boldsymbol{x},U_1,...U_l,\boldsymbol{P_1},..\boldsymbol{P_l})\, dU_1..dU_l d\boldsymbol{P_1}..d\boldsymbol{P_l} \qquad (2.1)$$

and impulse density functions $\boldsymbol{P_j}(t,x)$ as

$$\boldsymbol{P_j}(t,\boldsymbol{x}) = \int \boldsymbol{P_j}\, f(t,\boldsymbol{x},U_1,...U_l,P_1,..P_l)\, dU_1..dU_l d\boldsymbol{P_1}..d\boldsymbol{P_l} \qquad (2.2)$$

That allows define e-space velocities $\boldsymbol{v_j}(t,x)$ of densities $U_j(t,x)$ as

$$U_j(t,\boldsymbol{x})\boldsymbol{v_j}(t,\boldsymbol{x}) = \boldsymbol{P_j}(t,\boldsymbol{x}) \qquad (2.3)$$

Densities $U_j(t,x)$ and impulses $\boldsymbol{P_j}(t,x)$ are determined as mean values of aggregates of corresponding variables of separate e-particles with coordinates $x$. Functions $U_j(t,x)$ can describe macro densities of Investment and Loans, Assets and Debts and so on. To describe evolution of macro densities like Capital or Assets and etc., let remind that they are composed (Eq. 2.1-2.3) by corresponding variables of e-particles. However Assets of e-particle $l$ at point $x$ are determined by Buy or Sell transactions of Assets with e-particles at any points $y$ on e-space. Thus macro densities depend on transactions between e-particles. To describe evolution of macro densities let introduce macro transactions on e-space.

*2.2. Macro transactions on economic space*

To change amount of Assets e-particle should Buy or Sell them. Any e-particle at point *x* may carry out transactions with e-particles at any point *y* on e-space. Assets of e-particles at point *x* define macro Assets as function of *x* Eq.(2.1). Values of macro variables like Assets, Investment or Credits and etc. are defined during certain time interval. For example macro Credits at moment *t* determine Credits provided during certain time term *T* that may be equal day, quarter, year and etc. Thus any variable at time *t* is determined by factor *T* that indicates time term of accumulation of that variable. The same parameter *T* defines duration of transactions. Let further treat any transactions as rate or speed of change of corresponding variable. For example let treat transactions by Credits at moment *t* as total Credits provided during time term *dt*.

Transactions between agents are the only tools that implement economic and financial processes. In his Nobel Lecture [11] indicates that: "Direct interdependence between two processes arises whenever the output of one becomes an input of the other: coal, the output of the coal mining industry, is an input of the electric power generating sector". Let call variables of two e-particles as *mutual* if "the output of one becomes an input of the other". For example, Credits as output of Banks are *mutual* to Loans as input of Borrowers. Assets as output of Investors are *mutual* to Liabilities as input of Debtors. Any exchange between e-particles by *mutual* variables is carried out by corresponding transactions. Allocation of economic agents by their coordinates on e-space allows replace Leontief's specification of economics and macro-finance by industries with mapping economics on e-space. Thus we replace transactions between industries - inter-industry tables - with transactions between points *x* and *y* on e-space.

Let present transactions between e-particles on e-space in a more formal way. Let call that transactions between e-particle *1* at point *x* and e-particle *2* at point *y* determine economic or financial field $a_{1,2}(x,y)$ as exchange of variables $B_{out}(1,x)$ and $B_{in}(2,y)$ at moment *t* during time term *dt*. Let $a_{1,2}(x,y)$ be equal output variable $B_{out}(1,x)$ from e-particle *1* to e-particle *2* and equal input of variable $B_{in}(2,y)$ of e-particle *2* from e-particle *1* at moment *t*. So, $a_{1,2}(x,y)$ describes rate of change of variable $B_{out}(1,x)$ of e-particle *1* at point *x* due to exchange with e-particle *2* at point *y*. The same time $a_{1,2}(x,y)$ describes rate of change of variable $B_{in}(2,y)$ of e-particle *2* at point *y* due to exchange with e-particle *1*. Thus variable $B_{out}(1,x)$ of e-particle *1* at point *x* changes due to action of field $a_{1,2}(x,y)$ with all e-particles at point *y* as:

$$dB_{out}(1,x) = \sum_i a_{1,i}(x,y)\, dt \qquad (3.1)$$

and vice versa

$$dB_{in}(2,y) = \sum_i a_{i,2}(x,y)\, dt \qquad (3.2)$$

For example Credits-Loans field may describe Credits (output) from e-particle *1* to e-particle *2*. For such a case $B_{in}(2)$ equals Loans received by e-particle *2* and $B_{out}(1)$ equals Credits issued by e-particle *1* during certain time term *T*. Sum of field over all input e-particles equals speed of change of output variable $B_{out}(1)$ of economic particle *1*. Let assume that all extensive variables of economic particles can be presented as pairs of *mutual* economic or financial variables or can be describes by *mutual* variables. Otherwise there should be macro variables that don't depend on any economic or financial transactions, don't depend on Markets, Investment and etc. We assume that any economic or financial variable of e-particles depends of certain transactions between e-particles. For example Value of e-particle (Value of Corporation or Bank) don't take part in transactions but is determined by market transactions that define price of stocks of corresponding Bank or by variables like Assets and Liabilities, Credits and Loans, Sales and Purchases, Mergers and Acquisitions and etc. Let assume that all extensive variables can be described by Eq.(3.1, 3.2) or through other *mutual* variables. Let assume that economic or financial fields describe dynamics of all extensive variables of e-particles and hence describe evolution of macroeconomics and finance.

Now let describe transition from description of transactions between e-particle to macro transactions. Let assume that transactions between e-particles at point *x* and e-particles at point *y* are determined by exchange by *mutual* variables like Assets and Liabilities, Credits and Loans, Buy and Sell, and etc. Different transactions describe exchange by different *mutual* variables. For example *Assets-Liabilities (al)* transactions at time *t* describe a case when e-particle "one" at point *x* during time *dt* Invest (output) into Assets of amount *al* of e-particle "two" at point *y* and e-particle "two" at point *y* at time *t* during time *dt* receives Investment (input) that increase it's *Liabilities* on amount *al* in front of e-particle "one" at point *x*. Let give formal definition of economic and financial fields based on example of *Assets-Liabilities* transactions.

Let assume that macroeconomics is under action of *n* major risks and each e-particle on e-space $R^n$ at moment *t* is described by coordinates $x=(x_1,...x_n)$ and velocities $v=(v_1,...v_n)$. Let assume that at moment *t* there are $N(x)$ e-particles at point *x* and $N(y)$ e-particles at point *y*. Let state that velocities of e-particles at point *x* equal $v=(v_1,...v_{N(x)})$. Let state that at moment *t* $N(x)$ e-particles at point *x* carry Assets-Liabilities transactions $al_{i,j}(x,y)$ with e-particles $N(y)$ at point *y*. In other words, if e-particle *i* at moment *t* at point *x* allocates it's Assets equal $al_{i,j}(x,y)$ at e-particle *j* at point *y* then e-particle particle *j* at point *y* at moment *t* increases it's Liabilities by $al_{i,j}(x,y)$ in front of e-particle *i*. Let assume that all e-particles on e-space are "independent" and thus sum by *i* of Assets-Liabilities transactions $al_{i,j}(x,y)$ at point *x* on e-space $R^n$ at time *t* during *dt* equal rise of Liabilities $l_j(x,y)$ of e-particle *j* at point *y* in front of all e-particles at point *x* at moment *t*

$$l_j(x,y) = \sum_i al_{ij}(x,y) = a_j(x,y); \quad i = 1,...N(x); \quad j = 1,...N(y)$$

and equal rise $a_j(x,y)$ of Assets at moment $t$ during $dt$ of all e-particles at point $x$ allocated at e-particle $j$ at point $y$. Sum by $j$ of transactions $al_{i,j}(x,y)$ at point $y$ on e-space $R^n$ equals rise $a_i(x,y)$ of Assets of e-particle $i$ at point $x$ allocated at all e-particles at point $y$

$$a_i(x,y) = \sum_j al_{ij}(x,y) = l_i(x,y) \; ; \; i = 1, \ldots N(x) \; ; j = 1, \ldots N(y)$$

and equals rise of Liabilities of all e-particles at point $y$ in front of e-particle $i$ at point $x$. Let define transactions $al(x,y)$ between points $x$ and $y$ as

$$al(x,y) = \sum_{ij} al_{ij}(x,y); \quad i = 1, \ldots N(x); i = 1, \ldots N(y) \qquad (4.1)$$

$al(x,y)$ equals growth of Assets of all e-particles at point $x$ that are allocated at e-particles at point $y$ at moment $t$ and equals rise of Liabilities of all e-particles at point $y$ in front of all e-particles at point $x$ at moment $t$. Eq.(4.1) are very similar to Leontief's framework: we replace Leontief's output from one industry by output from all e-particles at point $x$ and input by second industry by input of all e-particles at point $y$. Transactions (4.1) between two points on e-space are random due to random character of deals between e-particles.

To introduce transactions as a regular functions and to derive equations that describe evolution of regular transactions on e-space let introduce analog of "transaction's impulse" alike to Eq.(1.1, 1.2) and [9]. Reasons to introduce transactions impulse are same as for impulses Eq.(1.1; 1.2). Velocities are non-additive and sum of agent's velocities don't define velocity of any group of agents. Impulses (4.2; 4.3) are additive and thus allow define impulse $P_j$ of fields by Eq.(5.2; 5.3). To do that let define additive variables $p_X$ and $p_Y$ that describe flux of Assets by e-particles along $x$ and $y$ axes. For Assets-Liabilities transactions $al$ let define *impulses* $p = (p_X, p_Y)$ alike Eq.(1.1; 1.2):

$$\boldsymbol{p}_X = \sum_{i,j} al_{ij} \boldsymbol{v}_i \; ; \quad i = 1, \ldots N(x); j = 1, \ldots N(y) \qquad (4.2)$$

$$\boldsymbol{p}_Y = \sum_{i,j} al_{ij} \boldsymbol{v}_j \; ; \quad i = 1, \ldots N(x); j = 1, \ldots N(y) \qquad (4.3)$$

Assets-Liabilities transactions $al(t,x,y)$ (4.1) and "impulses" (4.2, 4.3) take random values due to random properties of e-particles variables. To obtain regular functions let apply averaging procedures. Let introduce distribution function $f = f(t, z=(x,y); al, p=(p_X,p_Y))$ on $2n$-dimensional e-space $R^{2n}$ that determine probability to observe Assets-Liabilities field $al$ at point $z=(x, y)$ with impulses $p = (p_X, p_Y)$ at time $t$. Averaging of Assets-Liabilities transactions and their "impulses" by distribution function $f$ determine "mean" continuous "*media*" or hydrodynamic-like approximation of transactions as functions of $z=(x,y)$. Let call hydrodynamic-like approximations of macro transactions as macro fields. Assets-Liabilities field $AL(z=(x,y))$ and "impulses" $P=(P_X,P_Y)$ take form:

$$AL(t, z = (x,y)) = \int al \, f(t, x, y; al, \boldsymbol{p}_X, \boldsymbol{p}_Y) \, dal \, d\boldsymbol{p}_X \, d\boldsymbol{p}_Y \qquad (5.1)$$

$$\boldsymbol{P}_X(t, z = (x,y)) = \int \boldsymbol{p}_X \, f(t, x, y; al, \boldsymbol{p}_X, \boldsymbol{p}_X) \, dal d\boldsymbol{p}_X d\boldsymbol{p}_Y \qquad (5.2)$$

$$\boldsymbol{P}_Y(t, z = (x,y)) = \int \boldsymbol{p}_Y \, f(t, x, y; al, \boldsymbol{p}_Y, \boldsymbol{p}_Y) \, dal \, d\boldsymbol{p}_X d\boldsymbol{p}_Y \qquad (5.3)$$

That defines e-space velocity $v(t,z=(x,y))=(v_x(t,z),v_y(t,z))$ of field $AL(t, z)$:

$$P_X(t,z) = AL(t,z)v_X(t,z) \quad (5.4)$$

$$P_Y(t,z) = AL(t,z)v_Y(t,z) \quad (5.5)$$

Macroeconomic and financial fields may describe many important properties. Assets-Liabilities field *AL(t,z=(x,y))* describes distribution of rate of Investment made from point *x* to point *y* at moment *t*. Due to Eq.(2.1) integral of field *AL(x,y)* by variable *y* over e-space $R^n$ defines rate of Investment from point *x*. Integral of *AL(x,y)* by *x* over e-space $R^n$ determines speed of change of total Investment made at point *y* or change of Liabilities at point *y* in front of all e-particles of macroeconomics during time term *dt*. Integral of *AL(t,x,y)* by variables *x* and *y* on e-space describes function *A(t)* that equals rate of growth or decline of total Assets in macroeconomics or rate of change of total macro Liabilities during time term *dt*. To simplify the problem let treat transactions between e-particles as only tool for implementation of macro processes. Meanwhile Credits-Loans field *A(x,y)* define Credits landing from point *x* to point *y* at moment *t* during time term *dt*. Integral of *A(x,y)* by variable *y* over e-space defines speed of Credits allocation from point *x*. Integral of *A(x,y)* by *y* over e-space determines speed of Loans change at point *y*. Integral of *A(x,y)* by *x* and *y* over e-space defines total macro Credits *A(t)* provided at moment *t* or total macro Loans received during time term *dt*. Credits-Loans field *A(x,y)* can help determine position of maximum Creditors at point $x_C$ and position $y_B$ of maximum Borrowers of Credits and distance between them. Assets-Liabilities field can help define position of maximum Assets at point $x_A$ and position of maximum Liabilities at point $y_L$ and describe dynamics of distance between these points. These relations could be very important for macro modeling. Below we derive hydrodynamic-like equations to describe Credits-Loans field *A(x,y)*.

## 3. Hydrodynamic-Like Equations for Macro Fields

Let derive equations for case of Credits-Loans field *A(x,y)*. Let describe field *A(t,x,y)* alike to density functions on *2n*-dimensional e-space $R^{2n}$ [5-9]. Continuous Equations (6.1) and Equations of Motion (6.2) on *A(t,z)* take form:

$$\frac{\partial A}{\partial t} + div(\boldsymbol{v}A) = Q_1 \quad (6.1)$$

$$A\left[\frac{\partial \boldsymbol{v}}{\partial t} + (\boldsymbol{v} \cdot \nabla)\boldsymbol{v}\right] = \boldsymbol{Q}_2 \quad (6.2)$$

Left side of Eq.(6.1) describes flux of Credits *A(t,z)* through surface of unit volume on e-space $R^{2n}$ and $Q_1$ describes factors that change Credits-Loans field *A(t,z)*. Left side of Equation of Motion describes flux of impulse *P(t,z) = A(t,z)v(t,z)* through unit volume surface on e-space $R^{2n}$. Taking into account Continuity Equation then left side of Equation of Motion for simplicity takes form of Eq.(4.2). $\boldsymbol{Q}_2$ describes factors that change *A(t,z)* and velocity *v(t,z)*. Eq.(6.1, 6.2) describe change of left-side factors under the action of right-side factors. These equations become practical only when we define right-side factors $Q_1$ and $\boldsymbol{Q}_2$ taking into account macro processes.

Each field is determined by transactions between e-particles. To define factors $Q_1$ and $Q_2$ let take into account that for example Credits-Loans field depends on many factors like returns on Credits, returns on other Credits, alternative Investment and etc. Above we assume that fields determine evolution of any macro variable. If so fields should define factors in the right hand side of Eq.(6.1, 6.2). Let assume that fields $B_1(t,z)$, $B_2(t,z)$,… $B_m(t,z)$ that define factors $Q_1$ and $Q_2$ in the right hand side of Eq.(6.1, 6.2) on field $A(t,z)$ and are different from field $A(t,z)$. Let call such fields and their velocities those determine factors $Q_1$ and $Q_2$ as fields *conjugate* to field $A(t,z)$. Each macro field can depend on many *conjugate* macro fields. For simplicity let assume that *interactions* between macro fields are *local* so factors $Q_1$ and $Q_2$ are defined by differential operators on fields $B_1(t,z)$, $B_2(t,z)$,… $B_m(t,z)$. For example let assume that Credits-Loans macro field $A(t,z)$ that describe Credits provided from point *x* to point *y* depend on Payment-on-Credits field $B(t,z)$ that describe all payoffs from Borrowers at point *y* to Creditors at point *x*. Such assumptions simplifies interactions between macro fields and allows develop macro models in a self-consistent manner. Indeed simplest model of interaction between fields describes two *self-conjugate* fields: macro field $B(t,z)$ and it's velocities define right hand side factors $Q_1$ and $Q_2$ for equations (6.1, 6.2) on field $A(t,z)$ and vise versa. Such simple model allows obtain equations (6.1, 6.2) in a closed form and describe interactions between two macro fields. Below we show how simple model of interactions between two *self-conjugate* fields allows describe waves that have certain parallels to surface waves in physical fluids.

## 4. Surface-Like Waves of Macro Fields

Let study model of interactions between Credits-Loans field $A(t,z)$ and Payment-on-Credits field $B(t,z)$. Let assume that allocation of Credits is described by field $A(t,x,y)$ and defines amount of Credits provided from e-particles at point *x* (Creditors) to e-particles at point *y* (Borrowers) at moment *t* per time interval *dt*. Let assume that financial counter flow is described by Payment-on-Credits field $B(t,z)$ that defines payoff amount from Payers or Borrowers at point *y* to Payees or Creditors at point *x* per time interval *dt*. In other words – e-particles at point *x* provide Credits and e-particles at point *y* get Loans. Let simplify Credits-Loans deals and assume that Credits $A(t,x,y)$ provided at moment *t* are determined by counter flow payment defined by Payment-on-Credits field $B(t,z)$ at same moment *t* and vice versa. That allows use (6.1; 6.2) and describe two *conjugate* fields $A(t,x,y)$ and $B(t,x,y)$ their velocities $v(t,x,y)$ and $u(t,x,y)$ by hydrodynamic-like equations:

$$\frac{\partial A}{\partial t} + div(vA) = Q_{11} \; ; \; \frac{\partial B}{\partial t} + div(uB) = Q_{12} \qquad (7.1)$$

$$A\left[\frac{\partial v}{\partial t} + (v \cdot \nabla)v\right] = Q_{21} \; ; \; B\left[\frac{\partial u}{\partial t} + (u \cdot \nabla)u\right] = Q_{22} \qquad (7.2)$$

Let study simplest case on one-dimensional e-space *R* that describes transactions between e-particles under action of a single risk *X*. Fields *A* and *B* depend on two variables *x* and *y* and are determined on 2-dimensional e-space $R^2$. Macro fields are defined on macro domain determined by minimum and maximum risk grade values. Let take them as *0* and *X*. Thus macro domain on 2-dimensional space *(x,y)* is determined on square: *0<x<X* ; *0<y<X*. Here *0* – minimum or most secure and *X* - maximum or most risky coordinates of e-particles on e-space. Economic and financial perturbations can disturb borders of macro domain and cause disturbances of fields on borders. Such perturbations can induce macro waves *alike to* surface waves in physical fluids. Let study equations (7.1; 7.2) on simple macroeconomic square defined by

$$0 < x < X \ ; \ 0 < y < X \tag{7.3}$$

and describe perturbations of fields on border *y=X* of macro domain (7.3). This border defines maximum risk ratings – most risky position of *Borrowers* and *Payers*. As we assume macroeconomic domain in a steady state has boundary that is determined by relations *y=X*. Let assume that in steady state fields *A(t,z=(x,y))* and *B(t,z=(x,y))* equal zero for *y>X*. Let study possible waves that can be generated by perturbations of fields *A(t,z=(x,y))* and *B(t,z=(x,y))* and their velocities near surface *y=X*. Let define perturbations of this surface as *y=ξ(t,x)*. Interactions between two *conjugate* fields require that macro border *y= ξ(t,x)* should be common for both fields. Otherwise interaction between *A(t,z=(x,y))* and *B(t,z=(x,y))* will be violated. Time derivation of function *y=ξ(t,x)* determines *y*-velocities $v_y$ and $u_y$ of both e-fields on surface *y= ξ(t,x)* as:

$$\frac{\partial}{\partial t}\xi(t,x) = v_y(t,x,y=\xi(t,x)) = u_y(t,x,y=\xi(t,x)) \tag{7.4}$$

To derive macro surface-like wave equations let follow [14,15]. Let assume that potentials *φ* and *ψ* determine velocities **v** and **u** as:

$$\boldsymbol{v} = \nabla \varphi \ ; \ \boldsymbol{u} = \nabla \psi \tag{7.5}$$

and neglect nonlinear factors. Thus Equations of Motion (7.2) take form:

$$A_0 \frac{\partial \boldsymbol{v}}{\partial t} = Q_{21} \ ; \ B_0 \frac{\partial \boldsymbol{u}}{\partial t} = Q_{22} \tag{7.6}$$

Here $A_0$ and $B_0$ are constants. Let take Continuous Equations (7.1) for fields *A(t,**x**,**y**)* and *B(t,**x**,**y**)* as:

$$\frac{\partial A}{\partial t} + \nabla \cdot (A\boldsymbol{v}) = Q_{11} \ ; \ \frac{\partial B}{\partial t} + \nabla \cdot (B\boldsymbol{u}) = Q_{12} \tag{7.7}$$

To define factors $Q_{ij}$ let outline that fields *A(t,z=(x,y))* and *B(t,z=(x,y))* describe "*action-at-a-distance*" between Creditors at points *x* and Borrowers at point *y* on e-space. Let treat fields *A(t,z=(x,y))* and *B(t,z=(x,y))* as functions of variable *z=(x,y)* on 2-dimensional e-space. Let assume that factor $Q_{11}$ for Continuity Equation (7.7) on Credits-Loans field *A(t,z=(x,y))* is proportional to divergence of Payment-on-Credits velocity **u** and

$$Q_{11} = a_1 \nabla \cdot \boldsymbol{u} \ ; \ a_1 > 0 \tag{8.1.1}$$

Assumption (8.1.1) means that time derivative of field *A(t,z=(x,y))* grow up if divergence of **u** that define source of Payment-on-Credits velocity **u** flux from point is positive. That means growth of Payment-on-Credits flux and that attract Creditors to allocate more Loans. If divergence of **u** is negative then there is a runoff of Payment-on-Credits and that makes Creditors decrease rate of providing their Loans. Simply speaking Creditors prefer Borrowers who pay for Credits. To define factor $Q_{12}$ for Continuity Equation (7.7) on Payment-on-Credits $B(t,z=(x,y))$ assume that it is proportional to divergence of Credits-Loans velocity $v$:

$$Q_{12} = a_2 \nabla \cdot v \;;\; a_2 < 0 \qquad (8.1.2)$$

Assumption (8.1.2) means that time derivative of Payment-on-Credits $B(t,z=(x,y))$ decrease if flux of Credits-Loans defined by divergence of velocity $v$ positive. It occurs due to excess supply of Credits-Loans. If divergence of Credits-Loans velocity $v$ is negative that means decline of Credits available for Borrowers and that increase cost of Credits and Payment-on-Credits. Eq.(8.1.1; 8.1.2) present simple models of relations between Credits-Loans and Payment-on-Credits fields. It is reasonable to start description of complex processes with simple models.

Let study fields disturbances on macro border $y=X$. *Creditors* and *Payees* have coordinates $x$, $0<x<X$ and *Borrowers* and *Payers* have coordinates $y$, $0<y<X$. Let study influence of *Borrowers* positions at $y=X$ on Credits-Loans field $A(t,z=(x,y))$ and influence of *Payers* positions on border $y=X$ on Payment-on-Credits $B(t,z=(x,y))$ field. Disturbances of *Borrowers* and *Payers* at risk border $y=X$ may induce surface-like waves of Credits-Loans $A(t,z=(x,y))$ and Payment-on-Credits $B(t,z=(x,y))$ fields. To define $Q_{2i}$ factors for Equations of Motion (7.6) let assume that acceleration of Credits-Loans field velocity $v$ is proportional to gradient of Payment-on-Credits field $B(t,z=(x,y))$.

$$A_0 \frac{\partial}{\partial t} v = Q_{21} \sim b \nabla B \;;\; b > 0 \qquad (8.2.1)$$

Indeed, Credits-Loans flux (8.2.1) grows in the direction of higher Payment-on-Credits and thus $b>0$. Let assume that acceleration of Payment-on-Credits velocity $u$ is proportional to gradient of Credits-Loans field $A(t,z=(x,y))$.

$$B_0 \frac{\partial}{\partial t} u = Q_{22} \sim d \nabla A \;;\; d < 0 \qquad (8.2.2)$$

Payment-on-Credits decrease in the direction of higher Credits-Loans as positive gradient of Credits-Loans field $A(t,z=(x,y))$ results in excess Credits and funding and that decline cost of Credits and reduce Payment-on-Credits. Hence let take $d<0$. Thus we assume that Credits-Loans $A(t,z=(x,y))$ flow increase in the direction of higher Payment-on-Credits $B(t,z=(x,y))$. As well Payment-on-Credits flow (8.2.2) decrease in the direction of higher Credits-Loans determined by gradient of $A(t,z=(x,y))$. Now let introduce *"macro accelerations"* $h=(h_x,h_y)$ and $g=(g_x, g_y)$ that act on Credits-Loans $A(t,x,y)$ and Payment-on-Credits $B(t,x,y)$ fields respectively along axes $X$ and $Y$. Let assume that *"macro accelerations"* $h=(h_x,h_y)$ act on Creditors along axis $X$ and Borrowers along axis $Y$ and prevent them from excess risk and possible losses. Let propose that *"macro accelerations"* $g=(g_x, g_y)$ act on Recipients or Payees along risk axis $X$ and Payers-on-Credits along risk axis $Y$ and prevent them from surplus risk. Such *"macro accelerations"* $h$ and $g$ may protect

Creditors and Borrowers as well as Payees and Payers-on-Credits from excess risk. Let define "*macro accelerations*" $\mathbf{h}=(h_x,h_y)$ and $\mathbf{g}=(g_x, g_y)$ by potentials $H$ and $G$ as follows:

$$\frac{d}{dx}G = g_x \; ; \; \frac{d}{dy}G = g_y \; ; \; \frac{d}{dx}H = h_x \; ; \; \frac{d}{dy}H = h_y \tag{8.2.3}$$

Let take Equations of Motion (7.2) on fields $A(t,x,y)$ and $B(t,x,y)$ as:

$$A_0 \frac{\partial}{\partial t} v_x = Q_{21x} = -g_x B_0 + b\frac{\partial}{\partial x} B \; ; \; A_0 \frac{\partial}{\partial t} v_y = Q_{21y} = -g_y B_0 + b\frac{\partial}{\partial y} B \tag{8.3}$$

$$B_0 \frac{\partial}{\partial t} u_x = Q_{22x} = -h_x A_0 + d\frac{\partial}{\partial x} A \; ; \; B_0 \frac{\partial}{\partial t} u_y = Q_{22y} = -h_y A_0 + d\frac{\partial}{\partial y} A \tag{8.4}$$

Relations (7.5) allow present Eq.(8.3; 8.4) as

$$A_0 \frac{\partial}{\partial t}\frac{\partial}{\partial x} \varphi = -g_x B_0 + b\frac{\partial}{\partial x} B \; ; \; A_0 \frac{\partial}{\partial t}\frac{\partial}{\partial y} \varphi = -g_y B_0 + b\frac{\partial}{\partial y} B$$

$$B_0 \frac{\partial}{\partial t}\frac{\partial}{\partial x} \psi = -h_x A_0 + d\frac{\partial}{\partial x} A \; ; \; B_0 \frac{\partial}{\partial t}\frac{\partial}{\partial y} \psi = -h_y A_0 + d\frac{\partial}{\partial y} A$$

Then Credits-Loans field $A(t,x,y)$ and Payment-on-Credits $B(t,x,y)$ can be written as:

$$A(t,x,y) = A_0(1 + \frac{1}{d}H(x,y)) + \frac{B_0}{d}\frac{\partial}{\partial t} \psi(t,x,y) \tag{8.5.1}$$

$$B(t,x,y) = B_0(1 + \frac{1}{b}G(x,y)) + \frac{A_0}{b}\frac{\partial}{\partial t} \varphi(t,x,y) \tag{8.5.2}$$

Potentials $H$ and $G$ Eq.(8.2.3) describe model of steady state distributions of Credits-Loans field $A(t,x,y)$ and Payment-on-Credits $B(t,x,y)$ fields determined by Eq.(8.2.3; 8.5.1;8.5.2) on e-space for $\varphi=\psi=0$. We state that macro steady state distributions on e-space have nothing common with equilibrium states in statistical physics. We consider that macroeconomics and macro-finance have no equilibrium states in the meaning of statistical physics. We regard macroeconomics as strongly non-equilibrium system that is under random transitions from one steady state to another. Description of steady state distributions of macro fields on e-space establish important and tough problem that requires a lot of further studies. Description of macro fields steady distributions may help model macro policy that can manage transitions from one sustainable macro steady state to another. In this paper for simplicity we regard (8.5.3) with potentials $H$ and $G$ as linear functions with $\mathbf{h}$ and $\mathbf{g}$-*constant*.

$$H(x,y) = h_x x + h_y y \; ; \; G(x,y) = g_x x + g_y y \tag{8.5.3}$$

Eq.(8.5.1-3) present Credit-Loans and Payments-on-Credits fields $A(t,x,y)$ and $B(t,x,y)$ as:

$$A(t,x,y) = A_0(1 + \frac{1}{d}[h_x(x-X) + h_y(y-X)]) + \frac{B_0}{d}\frac{\partial}{\partial t} \psi(t,x,y) \tag{8.6.1}$$

$$B(t,x,y) = B_0(1 + \frac{1}{b}[g_x(x-X) + g_y(y-X)]) + \frac{A_0}{b}\frac{\partial}{\partial t} \varphi(t,x,y) \tag{8.6.2}$$

Eq.(8.5;8.6) present sample of distribution of Credits-Loans and of Payments-on-Credits over macro domain on e-space. Here $A_0$ and $B_0$ are constant values of fields $A(t,x,y)$ and $B(t,x,y)$ in steady state on macro domain at point with most risky coordinates $(X,X)$. $A_0$ define amount of Credits allocated at point $y=X$ by Creditors with coordinates $x=X$ at moment $t$ during time term $dt$. In other words $A_0$ – Loans received by most risky agents with coordinates $y=X$ from most risky

Creditors with coordinates *x=X*. Due to Eq.(8.2.1;8.2.2) *b>0* and *d<0* and $A_0$ is a minimum value of Credits-Loans field *A(t,x,y)* on macro domain on e-space. Borrowers on risk border *y=X* receive less Credits than those with coordinates *y<X*. $B_0$ – is a maximum value paid on Loans by Borrowers with most risky coordinates *y=X* to most risky Creditors with coordinates *x=X*. In other words - $B_0$ – is a maximum cost of Credits paid by most Risky Borrowers at point *y=X* to most risky Creditors at point *x=X*. Thus Eq.(8.6) is a simple model of common rule: risk cost money. In a steady state of macro domain (7.3) on border *y=X* fields *A* and *B* take form:

$$A(t,x,X) = A_0\left(1 + \frac{1}{d}h_x(x-X)\right); \quad B(t,x,X) = B_0(1 + \frac{1}{b}g_x(x-X)) \tag{8.7}$$

Relations (8.5; 8.6) on perturbation border *y= ξ(t,x)* take form:

$$A(t,x,\xi(t,x)) = A_0(1 + \frac{1}{d}[h_x(x-X) + h_y(\xi(t,x)-X)]) + \frac{B_0}{d}\frac{\partial}{\partial t}\psi(t,x,\xi(t,x))$$

$$B(t,x,\xi(t,x)) = B_0(1 + \frac{1}{b}[g_x(x-X) + g_y(\xi(t,x)-X)]) + \frac{A_0}{b}\frac{\partial}{\partial t}\varphi(t,x,\xi(t,x))$$

Let assume that fields *A(t,x,y)* and *B(t,x,y)* on surface *y= ξ(t,x)* take same values *A(t,x,X)* and *B(t,x,X)* as they have Eq.(8.7) in steady state on *y=X*. Hence:

$$A_0\left(1 + \frac{1}{d}[h_x(x-X) + h_y(\xi(t,x)-X)]\right) + \frac{B_0}{d}\frac{\partial}{\partial t}\psi(t,x,\xi(t,x)) = A_0\left(1 + \frac{1}{d}p_x(x-X)\right)$$

$$B_0\left(1 + \frac{1}{b}[g_x(x-X) + g_y(\xi(t,x)-X)]\right) + \frac{A_0}{b}\frac{\partial}{\partial t}\varphi(t,x,\xi(t,x)) = B_0(1 + \frac{1}{b}g_x(x-X))$$

Thus obtain:

$$\xi(t,x) - X = -\frac{B_0}{A_0 h_y}\frac{\partial}{\partial t}\psi(t,x,\xi(t,x)) = -\frac{A_0}{B_0 g_y}\frac{\partial}{\partial t}\varphi(t,x,\xi(t,x)) \tag{8.8}$$

Eq.(8.8) determine relations between $p_y$ and $g_y$

$$A_0^2 h_y = B_0^2 g_y$$

Eq.(7.4; 8.8) give:

$$\frac{\partial}{\partial t}\xi(t,x) = \frac{\partial}{\partial y}\psi = \frac{\partial}{\partial y}\varphi = -\frac{A_0}{B_0 g_y}\frac{\partial^2}{\partial t^2}\varphi(t,x,y=\xi(t,x)) \tag{8.9}$$

Eq.(8.9) describe constraints on potentials *φ* and *ψ* on surface *y=ξ(t,x)*. To derive equations on potentials *φ* and *ψ* let substitute Eq.(8.5; 8.6) into Continuity Equations (7.7; 8.1; 8.2) and neglect non-linear terms with potentials and "*macro accelerations*":

$$\left(A_0\frac{\partial^2}{\partial t^2} - a_2 b\Delta\right)\varphi = -bB_0\Delta\psi \; ; \; \left(B_0\frac{\partial^2}{\partial t^2} - a_1 d\Delta\right)\psi = -dA_0\Delta\varphi \; ; \; \Delta = \frac{\partial^2}{\partial x^2} + \frac{\partial^2}{\partial y^2}$$

Hence equations on potentials *φ* and *ψ* take form:

$$\left[\left(A_0\frac{\partial^2}{\partial t^2} - a_2 b \Delta\right)\left(B_0\frac{\partial^2}{\partial t^2} - a_1 d \Delta\right) - bdA_0 B_0 \Delta^2\right]\varphi = 0 \tag{9.1}$$

Let take potentials *φ* and *ψ* as:

$$\varphi = \psi = \cos(kx - \omega t)f(y - X) \; ; \; f(0) = 1 \tag{9.2}$$

Let take into account that perturbations $\xi(t,x)$ near steady boundary $y=X$ are small and hence relations (8.8) for (9.2) at $y=X$ give:

$$\frac{\partial}{\partial y} f(0) = \frac{A_0 \omega^2}{B_0 g_y} > 0 \tag{9.3}$$

and substitute (9.2) into (9.1). Then obtain equation on function $f(y)$ as ordinary differential equation of forth order:

$$\left(q_4 \frac{\partial^4}{\partial y^4} + q_2 \frac{\partial^2}{\partial y^2} + q_0\right) f(y) = 0 \tag{9.4}$$

$$q_4 = a_1 a_2 bd - bd B_0 A_0 \ ; \ q_2 = (A_0 \omega^2 - a_2 bk^2) a_1 d + (B_0 \omega^2 - a_1 dk^2) a_2 b + bd A_0 B_0 2k^2$$

$$q_0 = (a_2 bk^2 - A_0 \omega^2)(a_1 dk^2 - B_0 \omega^2) - bd A_0 B_0 k^4; \ q_1 = q_3 = 0$$

Characteristic equation of Eq.(9.4) take form

$$q_4 s^4 + q_2 s^2 + q_0 = 0 \tag{9.5}$$

Due to Eq.(8.1.1-8.2.2) $a_1 > 0; a_2 < 0; b > 0; d < 0$ thus $q_4 > 0; q_2 < 0; q_0 > 0$. Hence due to Vieta theorem all roots of eq.(9.5) are real - two positive $s_1>0; s_2>0$ and two negative $s_3 = -s_1 < 0; s_4 = -s_2 < 0$. Roots $s_i$ are determined by coefficients of Eq.(9.4) and Eq.(9.3) define constraints between frequency $\omega$ and wave number $k$ – dispersion relations that allow obtain group velocity of waves (9.2). Eq.(9.4) may has solutions due to roots $s_1,...s_4$ of it's characteristic polynomial (9.5) [19] and Eq.(9.2; 9.3):

$$f(y - X) = \sum_{i=1,4} \lambda_i \exp(s_i(y - X)) \ ; \ \sum_{i=1,4} \lambda_i = 1 \ ; \ \sum_{i=1,4} \lambda_i s_i = \frac{A_0 \omega^2}{B_0 g_y} > 0 \tag{9.6}$$

Simplest solution (9.6) for one real root $s>0$ gives potentials $\varphi$ and $\psi$ as:

$$\varphi = \psi = \cos(kx - \omega t) \exp(s(y - X)); \ f(y - X) = \exp(s(y - X)) \ ; \ \frac{A_0 \omega^2}{B_0 g_y} = s > 0 \tag{10.1}$$

Function $y=\xi(t,x)$ due to (5.9; 7.1) takes form:

$$\xi(t, x) = X - \frac{A_0 \omega}{B_0 g_y} \sin(kx - \omega t) = X - \sqrt{\frac{A_0 s}{B_0 g_y}} \sin(kx - \omega t) \tag{10.2}$$

Border $y=X$ define position of Borrowers for Credits-Loans field $A(t,x,y)$ and Payers-on-Credits for Payment-on-Credits field $B(t,x,y)$. Credits-Loans field $A(t,x,y)$ and Payment-on- Credits field $B(t,x,y)$ waves at stationary border $y=X$ take form:

$$A(t, x, X) = A_0 \left(1 + \frac{h_x}{d}(x - X)\right) + \frac{B_0 \omega}{d} \sin(kx - \omega t) \tag{10.3}$$

$$B(t, x, X) = B_0 \left(1 + \frac{g_x}{d}(x - X)\right) + \frac{A_0 \omega}{b} \sin(kx - \omega t) \tag{10.4}$$

Eq.(10.3) describes surface-like waves of Credits-Loans field $A(t,x,X)$ that reflect change of rate of Purchases carried by Borrowers with coordinate $y=X$ from Creditors with coordinates $x$. As well Eq.(10.4) describes change of rate of all Pay-off from Payers-on-Credits with coordinate $y=X$ to recipients with coordinates $x$. Integral of Credits-Loans field $A(t,x,X)$ (10.3) along border $y=X$ by over $(0,X)$ define field $A(t,X)$ at $y=X$ as function of time:

$$A(t,X) = A_0[X - \frac{h_x X^2}{2d}] + 2\frac{B_0 \omega}{dk} \sin\left(\omega t - k\frac{X}{2}\right) \sin\left(\frac{X}{2}k\right) \qquad (10.5)$$

Function $A(t,X)$ (10.5) describes total Credits provided to Borrowers with coordinates $y=X$ at risk border. In other words, (10.5) describe time oscillations with frequency $\omega$ of rate of Loans received by Borrowers with risk rating $y=X$ from all Creditors. Irregular time fluctuations of rate of Loans receive by Borrowers with risk rating at border $y=X$ may indicate action of random surface-like waves of Credits-Loans field $A(t,x,X)$ with random frequencies. Simplest solution Eq.(10.1) $s>0$ describe exponential dissipation of disturbances induced by surface-like waves inside macro domain $y<X$.

However there might be solutions that describe amplification of disturbances inside macro domain. Let take $s_1>0$ and $s_3=-s_1<0$ then for $\lambda_1 + \lambda_3 = 1$; $s_1(\lambda_1 - \lambda_3) > 0$ potentials:

$$\varphi = \psi = \cos(kx - \omega t)\,[\lambda_1 \exp(s_1(y-X)) + \lambda_3 \exp(-s_1(y-X))]$$

Taking into account Eq. (8.5; 8.6) Credits-Loans $A(t,x,X)$ and Payment-on-Credits fields $B(t,x,y)$ take form:

$$A(t,x,y) = A_0(1 + \frac{1}{d}[h_x(x-X) + h_y(y-X)]) + \frac{\omega B_0}{d} \sin(kx - \omega t)\,[\lambda_1 \exp(s_1(y-X)) + \lambda_3 \exp(-s_1(y-X))]$$

$$B(t,x,y) = B_0(1 + \frac{1}{b}[g_x(x-X) + g_y(y-X)]) + \frac{\omega A_0}{b} \sin(kx - \omega t)\,[\lambda_1 \exp(s_1(y-X)) + \lambda_3 \exp(-s_1(y-X))]$$

and Credits-Loans $A(t,x,X)$ and Payment-on-Credits $B(t,x,y)$ grow up as exponent for $(y-X)<0$

$$\sim \lambda_3 \frac{\omega B_0}{d} \sin(kx - \omega t)\,\exp(-s_1(y-X)) \qquad (10.6)$$

This example shows that small disturbances of rate of allocation Loans to Borrowers with maximum risk rating coordinate $y=X$ may induce exponentially growing (10.6) disturbances of rate of providing Credits from all Creditors with coordinate $x$ to secure Borrowers with coordinates $y<X$ far from risk border. Creditors run from risk and provide most of their Credits to secure Borrowers.

Thus Eq.(9.4) admits solutions that describe exponential amplification of Credits-Loans $A(t,x,y)$ and Payment-on-Credits $B(t,x,y)$ fields perturbations far from macro border $y=X$ that were induced by small perturbations on border $y=X$. Such disturbances inside macro domain far from maximum risk rating $X$ can induce macro instabilities and crises. Macro surface-like waves show hidden complexity of internal economic and financial processes. Studies of similar surface-like waves for different macro fields that describe various macro transactions between points of macro domain on economic space might be important for managing macro sustainability and require further investigations.

# 5. Conclusions

Fluctuations of Assets Prices, Credits and Loans, Demand and Supply, Investment and Profits are determined by numerous economic and financial factors. Description of any macro fluctuations is a very complex problem. Our paper presents example that economic space allows apply methods of theoretical physics to macro modeling and that reveals hidden complexity of macro processes. However we affirm that distinctions between macroeconomics and physics are so vital that almost any physical models are helpless. Indeed, even simplest model of macro transactions described by Eq.(9.1) has fourth order whereas most of mathematical physics is based on PDE of second order. Complexity of macro relations and processes that are described by macro fields interactions, hydrodynamic-like equations and wave equations on economic space leave few chances to derive similar results by "mainstream" economic theories. Diversities of wave processes that may occur in macroeconomics and finance proof importance of further studies of macro wave models on economic space. Economic space and macro model are based on well-known notions: economic agents, risk ratings, Leontief's framework. We just transform them a bit and make few assumptions on possible extension of current econometric and risk assessment methodologies and practice.

Usage of economic space for modeling macroeconomics and finance allows present transactions between agents with different risk ratings $x$ and $y$ as functions of variables $(x,y)$ on economic space. As example we present a simple model of mutual relations between Credit-Loans and Payment-on-Credits transactions and describe their interactions by hydrodynamic-like equations. We propose that macroeconomics admits steady states on economic space. We introduce potentials $H$ and $G$ that describe "*macro accelerations*" $h$ and $g$ and define model of steady state distributions of Credits-Loans field $A(t,x,y)$ and Payment-on-Credits $B(t,x,y)$ on economic space. We assume that macro steady state distributions have nothing common with equilibrium states in statistical physics. We consider that macroeconomics and finance have no equilibrium states in the meaning of statistical physics. We regard macroeconomics as strongly non-equilibrium system and it's evolution as random transitions from one steady state to another. Steady state distributions of macro fields on e-space establish important and tough problem. Modeling of macro steady states could describe dependence of macro variables on risk coordinates and model macroeconomic and financial policy that could manage transitions from one macro steady state to another. Modeling possible macro steady states on e-space requires a lot of further studies. In this paper for simplicity we regard potentials $H$ and $G$ as linear functions of $(x,y)$. Most secure and most risky agents establish risk borders of macro domain on economic space. Macro transactions may be disturbed if economic or financial shocks perturb agents on risk borders of macro domain. Such disturbances may propagate along borders *alike to* surface waves in fluids. We present surface-like waves that model case when small disturbances of Credits-Loans $A(t,x,y)$ and Payment-on-Credits $B(t,x,y)$ fields that describe rate of allocation of Loans to Borrowers with maximum risk rating coordinate $y=X$ may induce exponentially growing disturbances inside macro domain. Such waves describe model with exponentially growing rate of providing Credits from all Creditors with coordinate $x$ to secure Borrowers with coordinates $y<X$ far from risk border. In simple language it describe case with Creditors those run from risk and thus provide most of Credits to secure Borrowers. Such disturbances of macro fields inside macro domain far from maximum risk rating $X$ can induce macro instabilities and crises. Macro surface-like waves show hidden complexity of internal economic and financial processes. Studies of similar surface-like waves for different macro fields that describe various macro transactions between points of macro domain on economic space might be important for managing macro sustainability and require further investigations.

Up now it is impossible compare predictions of our models with econometric data. To do that, we propose develop risk assessment methods, econometric statistics and observations. We hope that such development may improve quality of economic and financial modeling and will assist macro policy-making. We hope that further development of our models will be useful for modeling macroeconomics and finance, Assets pricing, Volatility and etc.